\documentclass[a4paper]{jpconf}
\usepackage{graphicx,amsmath,cite}
\usepackage{iopams}

\usepackage{enumitem} 

\def \bl {\mbox{\boldmath{$\ell$}}}

\newcommand{\beq}{\begin{eqnarray}}
\newcommand{\eeq}{\end{eqnarray}}
\newcommand{\be}{\begin{equation}}
\newcommand{\ee}{\end{equation}}
\newcommand{\beqn}{\begin{eqnarray}}
\newcommand{\eeqn}{\end{eqnarray}}

\def\d{\mathrm{d}}

\begin{document}

\title{On type II(D) Einstein spacetimes in six dimensions}

\author{David Koko\v ska$^{1,2}$, Marcello Ortaggio$^1$}

\address{$^1$ Institute of Mathematics, Czech Academy of Sciences, \newline \v Zitn\' a 25, 115 67 Prague 1, Czech Republic}
\address{$^2$ Institute of Theoretical Physics, Faculty of Mathematics and Physics, \newline
 Charles University, V Hole\v{s}ovi\v{c}k\'{a}ch 2, 180 00 Prague 8, Czech Republic}

\ead{david.kokoska@matfyz.cz, ortaggio(at)math(dot)cas(dot)cz}

\begin{abstract}
After a concise overview of Einstein spacetimes of type II (or more special) in four and five dimensions, we summarize recent results in the six-dimensional case. We assume the optical matrix to be non-degenerate and ``generic'', and the Weyl tensor to fall off sufficiently rapidly at infinity. As it turns out, the most general metric is characterized by one discrete (normalized) and three continuous parameters, is of type~D and belongs to the Kerr-Schild class. Its relation to the previously known Kerr-(A)dS and Kerr-NUT-(A)dS metrics is clarified. 

\end{abstract}

\section{Introduction}

The celebrated Kerr solution was originally found ``as an example of algebraically special metrics'' (specifically, of type~D) \cite{Kerr63} in the sense of the Petrov classification of the Weyl tensor \cite{petrov}. The class of algebraically special metrics has been subsequently shown to include numerous exact solutions describing, among others, accelerating black holes with charge, a NUT parameter, and a cosmological constant, as well as wave-like geometries (we refer to \cite{Stephanibook,GriPodbook} for reviews and numerous references). In particular, the full family of type~D vacua is known \cite{Carter68pla,Carter68cmp,Kinnersley69,Debever71,Carter73,Plebanski75_2,PleDem76,GarciaD84,DebKamMcL84}.

For Lorentzian geometries in $n>4$ dimensions, an extension of the Petrov classification has been put forward in \cite{Coleyetal04}. Accordingly, higher-dimensional extensions of the Schwarzschild and Kerr(-NUT-AdS) black holes \cite{Tangherlini63,MyePer86,Chakrabarti86,HawHunTay99,Gibbonsetal04_jgp,CheLuPop06} are also of type~D \cite{Pravdaetal04,ColPel06,Hamamotoetal07,PraPraOrt07,OrtPraPra09}. 
However, although various properties of $n>4$ type~II/D metrics in higher dimensions have been elucidated \cite{PraPraOrt07,Durkee09,DurRea09,ParWyl11,Ortaggioetal12,OrtPraPra13,ReaGraTur13,deFGodRea15,Wylleman15,deFGodRea16,Ortaggio17,OrtPraPra18,TinPra19,Taghavi-Chabert22}, the full type~II/D class is so far known only for $n=5$ \cite{deFGodRea15,Wylleman15,deFGodRea16}. Under certain assumptions, a partial classification has been obtained also for $n=6$ for the case of a vanishing cosmological constant $\lambda$ \cite{Ortaggio17}. Recently, we have extended the results of \cite{Ortaggio17} (at least ``generically'', as specified below) to include an arbitrary $\lambda$. We note there is a qualitative difference in the structure of the Weyl tensor in $n=4,5$ as compared to $n=6$ (or any higher dimensions), for   the purely spatial components $C_{ijkl}$ bring in new ``degrees of freedom'' when $n>5$ \cite{PraPraOrt07,OrtPra14}. Furthermore, the case $n=6$ deserves attention also because only even dimensions allow for twisting Weyl aligned null directions (WANDs) with zero shear \cite{OrtPraPra07}, which is of interest in connection with higher-dimensional formulations of the Goldberg-Sachs theorem \cite{Pravdaetal04,OrtPraPra09,DurRea09,OrtPraPra09b,Ortaggioetal12,OrtPraPra13,OrtPraPra18,TinPra19,Taghavi-Chabert22}.

It is the purpose of the present contribution to summarize the recent results of \cite{KokOrt25} for the case $n=6$. However, to put the discussion in a broader context, we begin by briefly recalling the basic facts established in lower dimensions $n=4,5$ (section~\ref{sec_n=4,5}). Subsequently, we outline the main findings of 
\cite{KokOrt25} (section~\ref{sec_n=6}).

Let us first state the assumptions that will be understood throughout the paper. We will consider $n$-dimensional ($n\ge4$) spacetimes that are Einstein, i.e.,\footnote{Often a rescaled cosmological constant $\Lambda$ is used in the literature, related to $\lambda$ by $\lambda=\frac{2\Lambda}{(n-1)(n-2)}$.} 
\be
	R_{ab}=(n-1)\lambda g_{ab} , \qquad \lambda \equiv \dfrac{R}{n(n-1)} ,
	\label{einst}
\ee
and further obey the following assumptions:

\begin{enumerate}[series=mainlist] 

\item\label{ass1} The Weyl type is~II or more special in the classification of \cite{Coleyetal04}, which can be expressed as \cite{Ortaggio09}
\be
	\ell_{[e}C_{a]b[cd}\ell_{f]}\ell^b=0 ,
	\label{type_II}
\ee
such that the null vector field $\bl$ is a multiple Weyl aligned null direction (mWAND).

\item\label{ass2} The $(n-2)\times(n-2)$ optical matrix associated to $\bl$, defined by $L_{ij} \equiv \ell_{a;b}m^{a}_{(i)}m^{b}_{(j)}$ ($\mbox{\boldmath{$m$}}_{(i)}$ are $n-2$ orthonormal spacelike vectors orthogonal to $\bl$) \cite{Pravdaetal04,OrtPraPra07}, is {\em non-degenerate}, i.e., 
\begin{align}
\label{DetofL}
    \det L \neq 0.
\end{align} 

This assumption is typically relevant to black hole solutions and to asymptotically simple spacetimes (by contrast, black strings correspond to a degenerate optical matrix -- cf. \cite{OrtPraPra09,OrtPraPra09b,OrtPraPra11,ReaGraTur13} for related comments).

\end{enumerate}

Taking an affine parameter $r$ along $\bl$ as one of the spacetime coordinates, from now on we will have
\be
 \bl = \partial_r .
\ee

\section{Review in lower dimensions $n=4,5$}

\label{sec_n=4,5}

\subsection{Four dimensions}

For $n=4$, assumption~(\ref{ass1}) is equivalent to requiring the Petrov type to be~II or more special. The Goldberg-Sachs theorem \cite{GolSac62} then ensures that $\bl$ is geodesic and shearfree. Next, assumption~(\ref{ass2}) rules out the Kundt class \cite{Kundt61,Kundt62}, so that thanks to the Sachs equation one can write the optical matrix as \cite{Sachs61,NP}
\begin{align}
\label{Lin4D}
    L =\dfrac{1}{r^2+y^2}\begin{bmatrix} r & y \\ -y & r \end{bmatrix} .
\end{align}
The spacetime function $y$ is independent of $r$ and is proportional to the twist of $\bl$, and the special case $y=0$ defines Robinson-Trautman spacetimes \cite{RobTra62}. All special Petrov types II, D, III, and N can occur, and the most general solution, which contains integration functions, is not known \cite{Stephanibook,GriPodbook}. However, the type~D class has been fully integrated and contains only integration constants 
\cite{Kinnersley69,Debever71,Carter73,Plebanski75_2,PleDem76,GarciaD84,DebKamMcL84} (see \cite{GriPodbook} for more references). It notably includes the Kerr black hole \cite{Kerr63} and its extension with a cosmological constant \cite{Carter68pla,Carter68cmp,Carter73}.

In contrast to the $n=4$ case, non-geodesic mWANDs may exist when $n\ge5$ \cite{PraPraOrt07,Ortaggio07,Durkee09,GodRea09}. However, they are always accompanied by a geodesic one \cite{DurRea09}. Therefore, in what follows we can again assume $\bl$ to be geodesic, without losing generality. We further note that assumption~(\ref{ass2}) excludes the types~III and N in more than four dimensions \cite{Pravdaetal04,Kubicek_thesis,OrtPraPra18}. Additionally, mWANDs in higher dimensions need not to be shearfree \cite{MyePer86,FroSto03}, and they are necessarily shearing if they possess twist and $n$ is odd \cite{OrtPraPra07}.

\subsection{Five dimensions}

When $n=5$, using assumptions~(\ref{ass1}) and (\ref{ass2}) one arrives at \cite{Ortaggioetal12,deFGodRea15}
\be
L=
\operatorname{diag}\!\left(
\frac{1}{r^2+y^2}
\begin{bmatrix}
r & y \\
-y & r
\end{bmatrix},
\;\frac{1}{r}
\right) ,
\ee
where, as before, $y$ does not depend on $r$. The most general solution is specified by three parameters and is described by a Kerr-Schild metric of type~D \cite{deFGodRea15}. It consists of three invariant subclasses, depending on whether $y$ is a function, a non-zero constant, or zero:
\begin{enumerate}[label=\alph*)]

	\item $\d y\neq0$: the metric is locally isometric to the solution of \cite{MyePer86,HawHunTay99} with two unequal spins (one of which can be zero) or an analytic continuation thereof, except for a special limiting case (in which the second mWAND becomes Kundt, cf.~\cite{deFGodRea15} for details).

	\item $\d y=0\neq y$: the metric is locally isometric to the solution of \cite{MyePer86,HawHunTay99} with two equal, non-zero spins (again up to analytic continuation), or to a metric of \cite{ManSte04} in a special subcase. 

   \item $y=0$: here $\bl$ becomes twistfree and therefore one has a Robinson-Trautman spacetime \cite{PodOrt06}, which in five dimension corresponds to the (generalized) static metric of \cite{Tangherlini63}. 

\end{enumerate}

The explicit form of the metric for all cases can be found in \cite{deFGodRea15}.

The case when $L$ has lower rank has been analyzed in \cite{ParWyl11,ReaGraTur13,Wylleman15,deFGodRea16} (see also \cite{DurRea09,Ortaggioetal12}). The types~II, III and N cannot occur when rank$(L)=1$ \cite{Pravdaetal04,Wylleman15}.

\section{Results in six dimensions}

\label{sec_n=6}

\subsection{Additional assumptions}

When $n=6$, a full classification has been achieved under two further assumptions. First:

\begin{enumerate}[resume*=mainlist]

\item\label{ass3} The spatial part of the Weyl tensor falls off ``fast enough'' far away along $r$, i.e.,
\begin{align}
\label{bw0WeylAsymBehaviour}
    C_{ijkm} = o(r^{-2}) .
\end{align}

\end{enumerate}

Assumption~(\ref{ass3}) is identically satisfied by algebraically special vacuum spacetimes in four dimensions \cite{Sachs61} and by five dimensional vacua under assumption~\eqref{ass1}, \eqref{ass2} \cite{OrtPra14,deFGodRea15}. In any higher dimension it allows for a partial extension of the Goldberg-Sachs theorem \cite{OrtPraPra09b}, which in turn for $n=6$ enables one to arrive at \cite{OrtPraPra09,OrtPraPra10,Ortaggio17}\footnote{The canonical forms of the $n=6$ optical matrix without assuming~\eqref{ass3} have been obtained in \cite{TinPra19}.}
\begin{align}
\label{Lin6D}
    L = \mathrm{diag}\left( \dfrac{1}{r^2+y_1^2}\begin{bmatrix} r & y_1 \\ -y_1 & r \end{bmatrix}, \dfrac{1}{r^2+y_2^2}\begin{bmatrix} r & y_2 \\ -y_2 & r  \end{bmatrix} \right) .
\end{align} 
The above fall-off~\eqref{bw0WeylAsymBehaviour} is required for asymptotically flat \cite{OrtPraPra09,GodRea12} or AdS spacetimes \cite{AshDas00}, and is automatically satisfied by Kerr-Schild spacetimes \cite{OrtPraPra09,MalPra11}.

Finally, our last assumption reads:

\begin{enumerate}[resume*=mainlist]

\item\label{ass2b} In addition to being non-degenerate, the optical matrix $L$ is generic, in the sense that $|y_1|\neq|y_2|$ and $\d y_1\neq0\neq\d y_2$.

\end{enumerate}

The genericity assumption~(\ref{ass2b}) provides us with a technical advantage, in that it enables us to employ the functions $y_1$ and $y_2$ as preferred coordinates. Nevertheless, it may be dropped, with the analysis of the corresponding ``non-generic'' branches presented elsewhere \cite{KokOrt26}. 

The full classification in the case $\lambda=0$ was obtained in \cite{Ortaggio17}, without imposing~(\ref{ass2b}).

\subsection{Summary of results}

\subsubsection{General metric}

The most general $n=6$ Einstein spacetime which obeys the above assumptions~\eqref{ass1}--\eqref{ass2b} can be locally written as \cite{KokOrt25}
\beqn
\label{CarterPlebanskiEF}
    & & \mathrm{d}s^2 = 2\mathrm{d}r\left[ \mathrm{d}u + \left( y_1^2+y_2^2 \right)\mathrm{d}\phi_1 + y_1^2y_2^2\mathrm{d}\phi_2 \right] + \left( r^2 + y_1^2 \right)\dfrac{y_2^2-y_1^2}{\mathcal{P}(y_1)}\mathrm{d}y_1^2 + \left( r^2 + y_2^2 \right)\dfrac{y_1^2-y_2^2}{\mathcal{P}(y_2)}\mathrm{d}y_2^2 \nonumber \\ 
		& & \qquad {}+ \dfrac{\mathcal{P}(y_1)}{(r^2+y_1^2)(y_2^2-y_1^2)}\left[ \mathrm{d}u + \left( y_2^2-r^2 \right)\mathrm{d}\phi_1 - r^2y_2^2\mathrm{d}\phi_2 \right]^2\nonumber \\ 
		& & \qquad {} + \dfrac{\mathcal{P}(y_2)}{(r^2+y_2^2)(y_1^2-y_2^2)}\left[ \mathrm{d}u + \left( y_1^2-r^2 \right)\mathrm{d}\phi_1 -r^2y_1^2\mathrm{d}\phi_2 \right]^2 \nonumber \\ 
		& & \qquad {}+\dfrac{\mathcal{Q}(r)}{(r^2+y_1^2)(r^2+y_2^2)}\left[\mathrm{d}u + \left( y_1^2+y_2^2 \right)\mathrm{d}\phi_1 + y_1^2y_2^2\mathrm{d}\phi_2 \right]^2 ,
\eeqn
with 
\be
  {\cal P}(s)\equiv\lambda s^6 + 2\hat{\mathcal{U}}^0s^4 - c_0s^2 - d_0 , \qquad {\cal Q}(r)\equiv\lambda r^6 -2\hat{\mathcal{U}}^0r^4 - c_0r^2 + \mu r + d_0 ,
\ee 
where $\hat{\mathcal{U}}^0$, $c_0$, $d_0$, and $\mu$ are integration constants (but only three are essential thanks to a scaling freedom).

Metrics~\eqref{CarterPlebanskiEF} are of constant curvature iff $\mu=0$, which demonstrates that they belong to the Kerr-Schild class \cite{OrtPraPra09,MalPra11}. A coordinate transformation presented in \cite{KokOrt25} reveals that they are locally isometric to a subfamily of the general Kerr-NUT-(A)dS family of \cite{CheLuPop06}. They are thus of type~D \cite{Hamamotoetal07}. In \cite{KokOrt25} we also briefly commented on the Kerr-Schild double copy \cite{MonOCoWhi14,BahLunWhi17,CarPenTro18} in spacetimes~\eqref{CarterPlebanskiEF}, the corresponding key findings dating back to \cite{CheLu08,Krtous07} (cf. \cite{MyePer86,Aliev07,Aliev07_2,FroKrtKub17,ChaKee23,OrtSri24} for related results).

\subsubsection{Special subclass: the doubly-spinning Kerr-(A)dS metric}

For suitable choices of the parameters $\hat{\mathcal{U}}^0$, $c_0$, and $d_0$, the function ${\cal P}(s)$ is fully factorizable and can be written as 
	\be
		{\cal P}(s)=(\lambda s^2+\epsilon)(s^2-a_1^2)(s^2-a_2^2), \qquad \epsilon=0,\pm1 ,
	\ee
where $\epsilon$, $a_1$, and $a_2$ correspond to combination of $\hat{\mathcal{U}}^0$, $c_0$, and $d_0$ (after also using the scaling freedom mentioned above). This implies ${\cal Q}(r)=-(\epsilon-\lambda r^2)(r^2+a_1^2)(r^2+a_2^2)+\mu r$.

As discussed in \cite{KokOrt25}, the case $\epsilon=1$ gives rise to the doubly spinning Kerr-(A)dS metric of \cite{Gibbonsetal04_jgp}, while the cases $\epsilon=0,-1$ correspond to the generalizations of \cite{MarPeo22_b,ChrConGra25} (see also \cite{Ortaggio17} when $\lambda=0$ and $\epsilon=-1$, and the earlier results with $n=4$\cite{KleMorVan98} and $n=5$ \cite{deFGodRea15} for arbitrary $\lambda$ and $\epsilon$). The line-element~(143,\cite{KokOrt25}) (see also Appendix~C of the same reference) encompasses the three branches in a unified way.

\ack

Supported by the Institute of Mathematics, Czech Academy of Sciences (RVO 67985840) and research grant GA\v CR~25-15544S.

\section*{References}

%

\providecommand{\newblock}{}

\end{document}